# The Galactic Habitable Zone and the Age Distribution of Complex Life in the Milky Way

**Charles H. Lineweaver,[1,2]\*  Yeshe Fenner,[3]\*  Brad K. Gibson[3]\***

[1]Department of Astrophysics, University of New South Wales, Sydney, NSW 2052, Australia.
[2]Australian Centre for Astrobiology, Macquarie University, NSW 2109, Australia.
[3]Centre for Astrophysics & Supercomputing, Swinburne University, Hawthorn, VIC 3122, Australia.
\*To whom correspondance should be addressed. E-mail: charley@bat.phys.unsw.edu.au (C.H.L.); yfenner@astro.swin.edu.au (Y.F.)

**We modeled the evolution of the Milky Way to trace the distribution in space and time of four prerequisites for complex life: the presence of a host star, enough heavy elements to form terrestrial planets, sufficient time for biological evolution and an environment free of life-extinguishing supernovae. We identified the Galactic habitable zone (GHZ) as an annular region between 7 and 9 kiloparsecs from the Galactic center that widens with time and is composed of stars that formed between 8 and 4 billion years ago. This GHZ yields an age distribution for the complex life that may inhabit our Galaxy. We found that 75% of the stars in the GHZ are older than the Sun.**

As we learn more about the Milky Way Galaxy, extrasolar planets and the evolution of life on Earth, qualitative discussions of the prerequisites for life in a Galactic context can become more quantitative (*1-3*). The Galactic habitable zone (GHZ) (*4*), analogous to the concept of the circumstellar habitable zone (*5*), is an annular region lying in the plane of the Galactic disk possessing the heavy elements necessary to form terrestrial planets and a sufficiently clement environment over several billion years to allow the biological evolution of complex multicellular life. In order to more quantitatively estimate the position, size and time evolution of the GHZ, we combined an updated model of the evolution of the Galaxy (*6*) with metallicity constraints derived from extrasolar planet data (*7*).

Of the factors that determine the location of the GHZ, the abundance of elements heavier than hydrogen and helium (metallicity) is particularly crucial because these elements are what terrestrial planets are composed of. The current metallicity of the Galaxy can be directly measured. However, modeling is needed to identify the metallicity distribution throughout the history of the Milky Way.

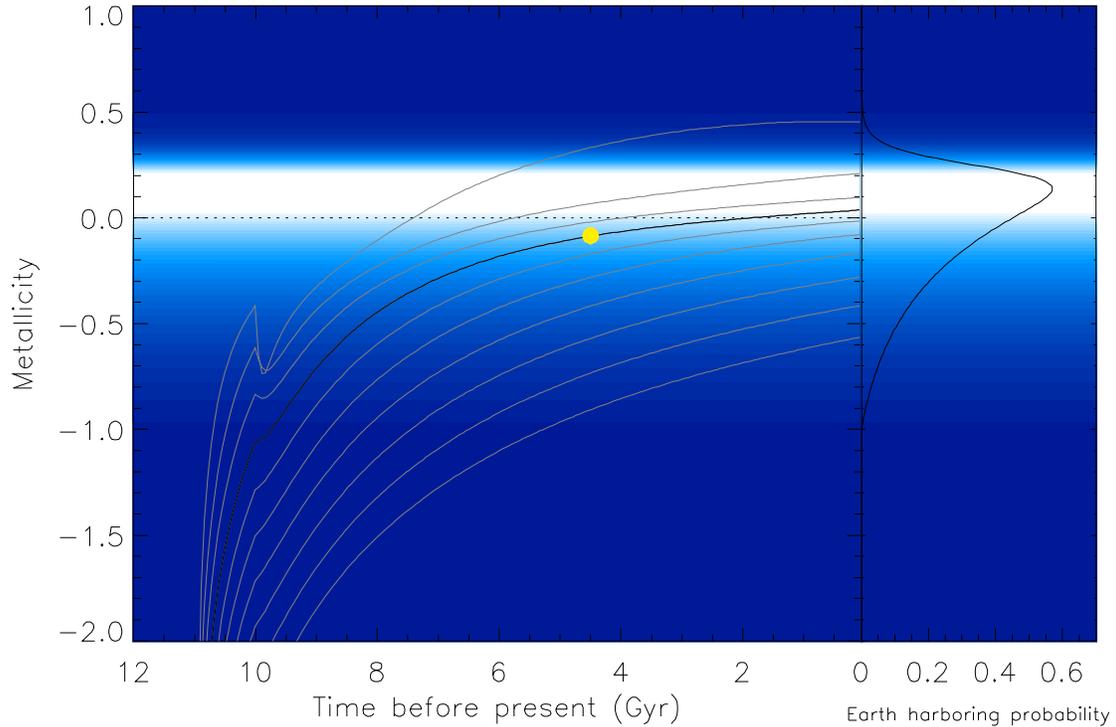

**Fig. 1.** The buildup of metals in our Galaxy as a function of time predicted by our simulations. Metallicities at different Galactocentric distances can be compared with the probability of harboring terrestrial planets as a function of the metallicity of the host star [**right**, see (7) for details]. Galactocentric distances from 2.5 kpc (upper curve) to 20.5 kpc (lower curve) are shown in 2 kpc increments. The yellow dot indicates the Sun's time of formation and Galactocentric distance of 8.5 kpc. The inner Galaxy accrues metals early and rapidly because of a high rate of star formation, whereas the most distant regions remain deficient in the metals needed to form terrestrial planets. The metallicity is the log of the ratio of the amount of iron to hydrogen in the stars relative to the Sun.

We simulated the formation of the Galaxy with the use of two overlapping episodes of accretion that correspond to the build-up of the halo and disk. The gas accretion rate falls off exponentially on a small [~1 Gyear (Gy)] time scale for the first phase and a longer time scale (~7 Gy) for the second phase. Although there is a 1 Gy delay between the onset of halo formation and the onset of thin disk formation, the formation of these two components overlaps in time. In our model, we monitor the creation of heavy elements and the exchange of matter between stars and gas. Model parameters have been chosen to reproduce the key observational constraints, namely, the radial distribution of stars, gas and metals; the metallicity distribution of nearby stars; and the chemical composition of the Sun (*6*).

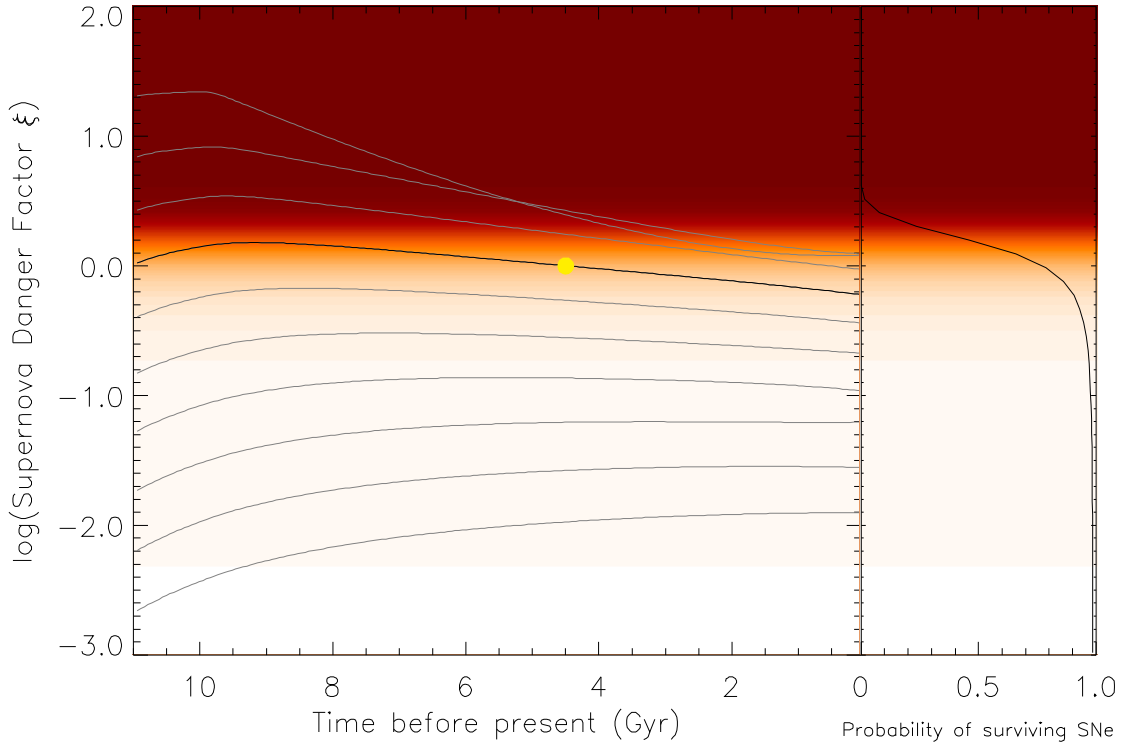

**Fig. 2.** The supernova danger factor, ξ, defined as the integral of the supernova rate from *t* to *t* + 4 billion years, in units of the Earth's ξ and plotted as a function of time for the same set of Galactocentric distances shown in Fig. 1. In the inner disk, the threat from supernovae (SNe) was once 20 times greater than that faced by the Earth. The inner disk began to run out of fuel for making stars within several billion years, causing both the star formation and supernova rates to decline.

Statistical analyses of extrasolar planets show a strong correlation between the presence of large close-orbiting massive planets and high metallicity of the host star (*8, 9*). Whether these planets slowly migrate in or are gravitationally perturbed into these close orbits, they may destroy Earth-mass planets as they pass through the circumstellar habitable zone. Thus, there is a Goldilocks zone of metallicity: with too little metallicity, Earth-mass planets are unable to form; with too much metallicity, giant planets destroy Earth-mass planets. A metallicity-dependent probability, $P_{metals}$, of harboring terrestrial planets (7) has been assigned to the space-time distribution of metals (Fig. 1).

Estimates of the GHZ need to include sufficient time to allow the biological evolution of complex life. On Earth this took ~ 4 Gy. Various arguments have been presented to explain why this duration should or should not be considered typical of the time scale for the evolution of life in general (*10-12*). Without making any assumption about how probable the evolution of complex life is, we assume that Earth's time scale is typical and adopt 4 ± 1 Gy as the characteristic time required for the evolution of complex life. We include this constraint as a probability $P_{evol}(t)$, defined as the cumulative integral of a normal distribution of mean 4 Gy and dispersion 1 Gy.

The deaths of massive stars (more than about eight times the mass of the Sun) produce supernovae that trigger blast waves and release cosmic rays, gamma-rays, and x-rays that can be fatal to life on nearby Earth-mass planets (*13*). However, there are large uncertainties concerning how robust organisms are to high radiation doses and climatic disturbances (*14*). The effects of nearby explosions depend sensitively on the thickness and composition of an atmosphere and on the density of intervening dust and gas. Massive but incomplete extinction of life may slow or speed up the evolution of complex life. In the face of these unknowns, we define the supernova danger factor $\xi(r,t)$ at Galactocentric distance $r$ and the time of a star's formation $t$, as the supernova rate integrated over the interval $t$ to $t + 4$ billion years (Fig. 2). We normalize relative to Earth. That is, we set the probability that complex life survives supernovae to be $P_{SN} = 0.5$ for $\xi$ equal to twice that of Earth (at a Galactocentric distance of 8.5 kpc), $P_{SN} \sim 0$ for $\xi$ equal to four times that of Earth, and $P_{SN} \sim 1$ for $\xi$ equal to half that of Earth (Fig. 2). This Earth normalization is somewhat arbitrary, because we have only imprecise notions of the vulnerability of terrestrial life to supernovae.

To describe the GHZ we define a probability $P_{GHZ}(r,t)$ as the product of four terms:

$$P_{GHZ} \; = \; SFR \; \times \; P_{metals} \; \times \; P_{evol} \; \times \; P_{SN} \qquad (1)$$

where $P_{metals}$, $P_{evol}$ and $P_{SN}$ are described above. Multiplication by the star formation rate *SFR*, expresses the fact that if there are more new stars in a given region, there are more potential homes for life. The stellar initial mass function used here is described in (15).

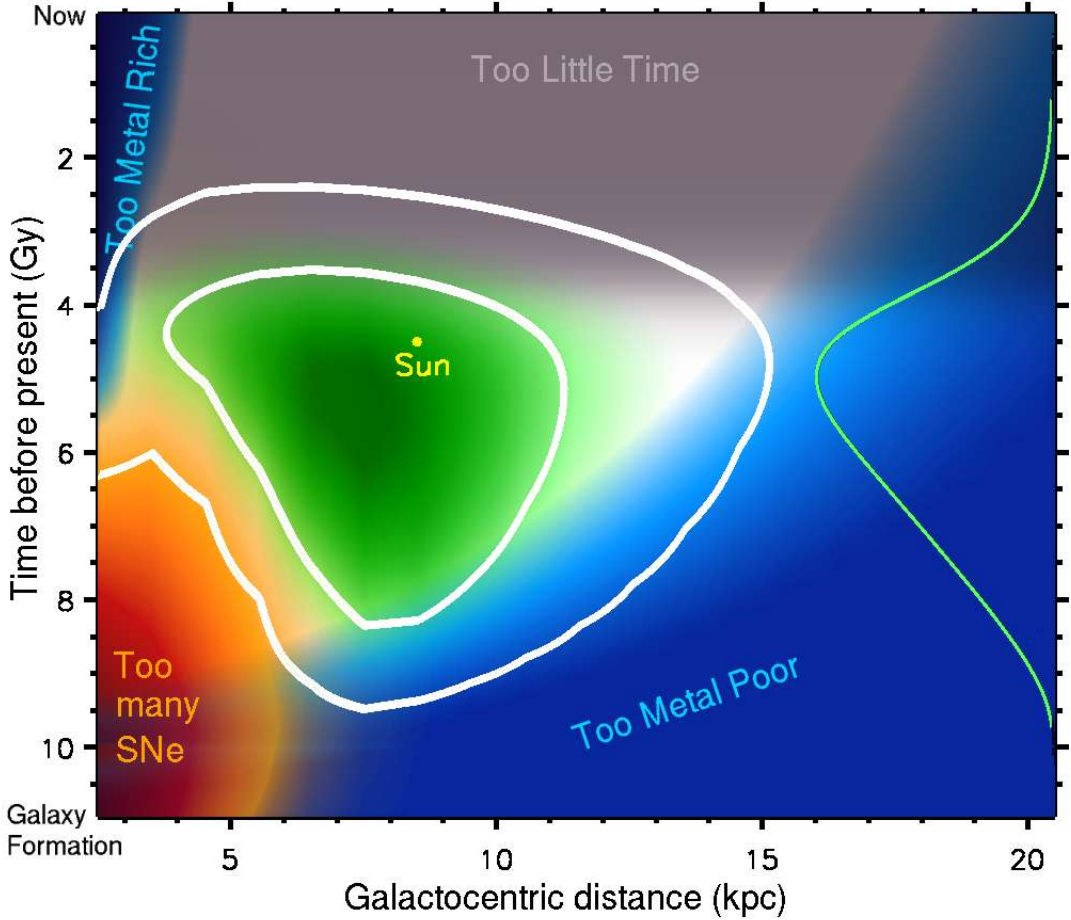

**Fig. 3.** The GHZ in the disk of the Milky Way based on the star formation rate, metallicity (blue), sufficient time for evolution (gray), and freedom from life-extinguishing supernova explosions (red). The white contours encompass 68% (inner) and 95% (outer) of the origins of stars with the highest potential to be harboring complex life today. The green line on the right is the age distribution of complex life and is obtained by integrating $P_{GHZ}(r, t)$ over *r*.

Although $P_{SN}$ is not independent of the *SFR*, the sign of the effect is opposite: with a higher *SFR*, the supernovae danger increases and therefore $P_{SN}$ decreases. With these inputs, $P_{GHZ}(r,t)$ is the relative number of potentially suitable planetary systems as a function of space and time. The regions encompassing 68% and 95% of these systems (Fig.

3) identify the GHZ. The 68% contour contains less than ~10% of the stars ever formed in the Milky Way.

Early intense star formation toward the inner Galaxy provided the heavy elements necessary for life, but the supernova frequency remained dangerously high there for several billion years. Poised between the crowded inner bulge and the barren outer Galaxy, a habitable zone emerged about 8 Gy ago (68% contour) that expanded with time as metallicity spread outward in the Galaxy and the supernovae rate decreased. By comparing the age distribution of the right of Fig. 3 to the origin of the Sun, we find that ~ 75% of the stars that harbor complex life in the Galaxy are older than the Sun and that their average age is ~ 1 Gy older than the Sun.

We do not assume that life or complex life is probable. The space-time distribution for the prerequisites of life determined here should be valid whether life is rare or common in the Galaxy. It is not surprising that the Sun lies within the GHZ, because we have used our local conditions as a template for finding similar systems. The Sun's values of $\xi$ and solar metallicity and Earth's biological time scale were used to help define the GHZ. However, independent information comes from extrasolar planet host metallicities, the supernova rate, and the metallicity evolution of the Galaxy. Our procedure is solar-centric but not to the extent of putting the Sun in the center of the GHZ.

The probability $P_{metals}$ will be improved when we learn more about extrasolar planets and the dependence of terrestrial planet formation on the metallicity of the prestellar nebula (*16*). The parameters used to set $P_{SN}$ and $P_{evol}$ have the virtues of being explicit and easily modified to accommodate new information; however they are somewhat simplistic. Assuming a 4 Gy typical time scale for life in the universe is highly speculative. However, if we are interested in all life (not just complex life) we can drop this assumption and exclude $P_{evol}$ from Eq. 1. The resulting GHZ (Fig. 4) resembles the GHZ of Fig. 3, but extends all the way to today. Integrating this GHZ to obtain the age distribution of life yields the result that ~30% of stars harboring life in the Galaxy are older than the Sun and that their average age is ~ 1 Gy younger than the Sun.

The Milky Way contains populations of stars belonging to an inner bulge component, a diffuse halo component, and a thick disk component; none of which appear as hospitable to life as the thin disk modeled here.  Low metallicity makes it unlikely that the halo and thick disk are home to many terrestrial planets. Conversely, the bulge suffers from a high density of stars, which results in a pervasive, intense radiation field (14), and more numerous close encounters between stars.

A previous analysis (4) based on the outward spread of a narrow range of metallicity, found that the GHZ has been slowly creeping outward.  We find that the GHZ remains centered at ~ 8 kpc but broadens with time (compare their figure 2 with our Fig. 3).

With the use of a chemical evolution model of the Galaxy to trace four plausible prerequisites of life, we have identified the space time regions most likely to harbor complex life.  This result depends on the assumption that the terrestrial time scale for biological evolution is representative of life elsewhere. These results will be directly tested within a few decades, when we are able to obtain a statistical distribution of terrestrial planets over a large range of Galactocentric distances.

These results can be improved by better modeling of the Galaxy and better understanding of the effects of various elements on the origin and evolution of life.  For example, the abundances of radioactive species such as $^{26}$Al play an important role in planet heating.  However, to first order the overall metallicity is a good gauge of the abundances of other elements; the abundance (relative to solar) of most elemental species usually falls within a factor of two of the iron abundance (relative to solar) in nearby stars with a wide range of metallicity.  Other factors that may play an important role should be visited in future studies.  These include the frequency of grazing impacts with molecular clouds,  the circularity of stellar orbits and their proximity to the corotation circle, and the effect of starbursts and an active Galactic nucleus in the early history of the most central regions of the Milky Way (17). If these additional constraints are used to further constrain the GHZ, it will encompass fewer stars, and the 10% quoted here will be recognized as an upper limit to the number of stars in the GHZ.

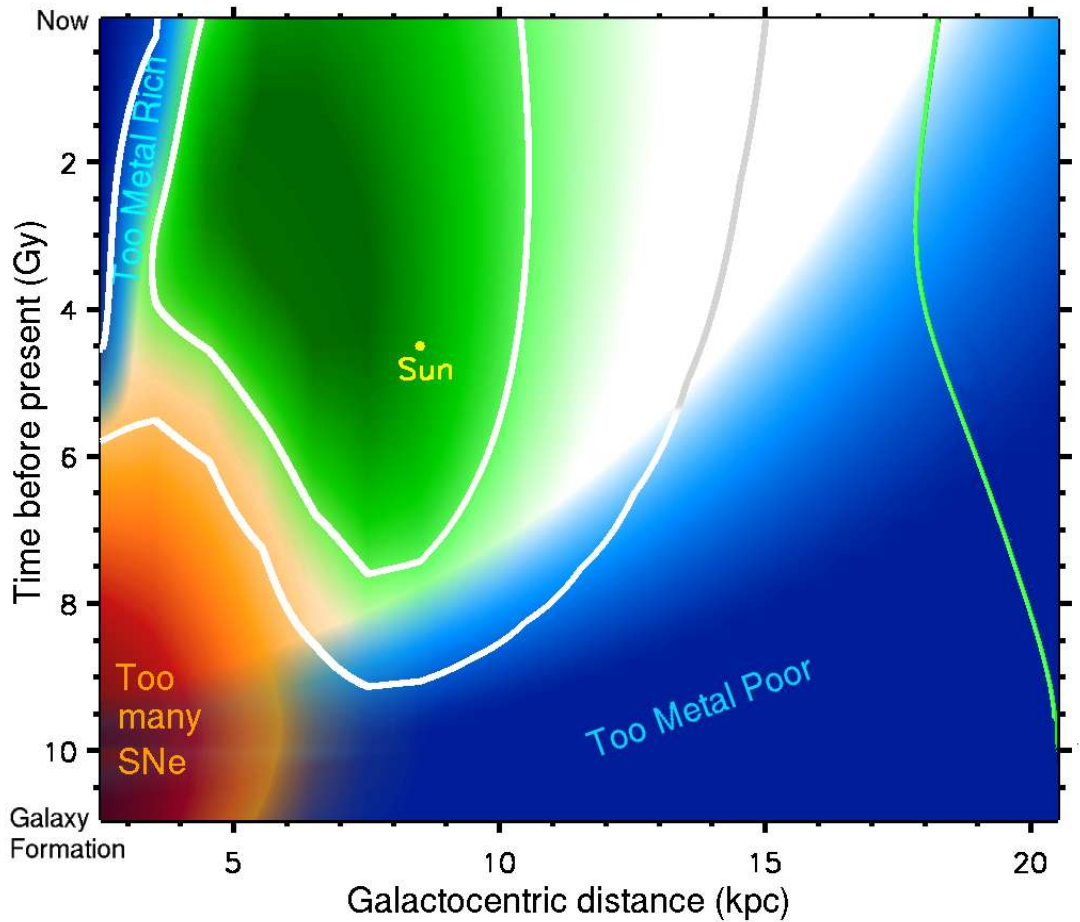

**Fig. 4.** The GHZ as in the previous figure but without requiring 4 ± 1 Gy for the evolution of complex life. Thus, the white contours here circumscribe the origins of stars with the highest potential to be currently harboring any life, not just complex life. The green line on the right is the resulting age distribution of life within the GHZ and is obtained by integrating $P_{GHZ}(r, t)$ over $r$.